# Desired Control of Mutually Delay-Coupled Diode Lasers near Phase-flip Transition Regimes

Pramod Kumar[*]

*Femtosecond Laser Laboratory, Department of Physical Sciences, Indian Institute of Science Education and Research (IISER) Mohali, Knowledge City, Sector 81, S.A.S. Nagar, Manauli-140306 Punjab, India*

**Abstract.** We investigate zero-lag synchronization (ZLS) between delay-coupled diode lasers system with mutual optical injection in a face-to-face configuration. We observed numerical evidence of such ZLS without using any relay element or mediating laser. In addition, simulation also demonstrate that this kind of robust ZLS occurs around the phase flip transition regimes where in-phase and anti-phase oscillations coexist due to delayed coupling induced modulation of the phase-amplitude coupling factor α. Our finding could be implemented in highly secured optical communication network as well as the understanding of the occurrence of such ZLS in the neural network functionality.

Keywords: Laser Diode, Delayed-coupling, Amplitude Death, Dynamical Instability.

PACS: 05.45.Xt, 42.55.Px, 05.45.-a

## INTRODUCTION

Mutually delay-coupled diode lasers system show a plethora of complex dynamical instability in the collective behavior. On the one hand, delay increases the dimensionality and hence the complexity but on the other hand delay can exert the optical phase shift in the functionality of the coupled lasers. When there is a time delay in the coupling then there are fascinating collective phenomena, such as phase-flip transitions, is observed. These transitions are characterized by the change of the locked dynamics from being in-phase to anti-phase or vice versa. The phase difference between the mutually coupled lasers undergoes a jump from 0 to $\pi$ as a function of coupling strength, $\eta$ and time-delay, $\tau$. However this phase-flip bifurcation does not occur abruptly at a particular combination of the control parameters $\eta$ and $\tau$. Instead, we find regimes around the phase-flip transition where multistability occurs. In general the form of interaction between the lasers together with many degrees of freedom such as phase, amplitude, frequency, polarization, and gain their intrinsic response provide sufficiently information about the ability of the system to synchronize or desynchronize the oscillation. It is quite probable that varieties of mechanisms are responsible for bringing synchrony at slow and fast time scales. Recent studies indicate that isochronous synchronization [6, 7] can be obtained by adding individual feedback to each laser [6] or inserting a relay laser [6], mirror [8], and multiple coupling delay time with well defined ratio between them. So the physical mechanism of phase locking and synchronization has been subject of controversial debate for many years.

In this paper, we address the question of how zero-lag synchronization can be achieved even in the presence of finite delay without any relay element. Zero-lag synchronization means that the two lasers' outputs synchronize with no time delay. The complex dynamical instabilities is investigated near the phase flip transition regimes as a function of coupled-cavity time delay $\tau$ and the optical injection strength $\eta$. We show that the phase flip transition does not occur abruptly at a particular value of coupling strength. Instead we find a coupling strength region around the phase flip transition where the co-existence of multi-attractors occurs.

## THEORETICAL MODEL

Here we present a simple face-to-face configuration of delay-coupled diode lasers as shown in figure 1. We have chosen semiconductor diode laser for our study, since they have proven to be excellent photonic test-bed to investigate the collective behavior of delay-coupled oscillator functionality. For numerical simulations, the time evolutions of the complex electric field $E_j(t)$ and the carrier density $N_j(t)$ (with the threshold value subtracted out) averaged spatially over the laser medium for each laser are modeled by [1, 2]. where η is the coupling parameter, $J$s are the injected current densities (with the threshold value subtracted out), $T$ is the ratio of the carrier lifetime to the photon lifetime, the delay time $\tau = L/c$ is the round-trip time taken by the light to cover the distance $L$ between the lasers, and α [3, 4, 5] is the line width enhancement factor of the diode lasers.

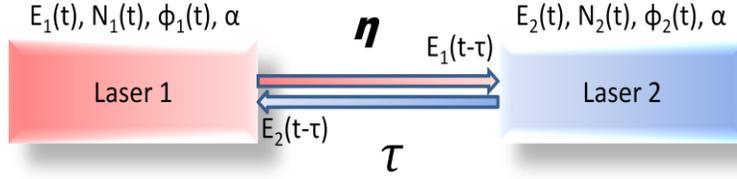

**FIGURE 1.** Scheme of two semiconductor diode lasers mutually delay-coupled in a face-to-face configuration.

$$\frac{dE_j(t)}{dt} = (1+i\alpha)N_j(t)E_j(t) + \eta E_k(t-\tau)e^{-i\omega_k\tau}, \ j \neq k,$$

$$T\frac{dN_j(t)}{dt} = J_j - N_j(t) - (1+2N_j(t))|E_j(t)|^2, \ j,k = 1,2.$$

The inclusion of time delay into the system is natural in the realistic consideration of finite transmission of interaction. To investigate the dynamics of the system we have performed numerical simulation of the above equations considering globally coupled system. In the simulation, we have used the forth-order Runge-Kutta method with discrete time steps of $\Delta t$ 0.001 [5]. At each run, the first $10^5$ time steps per oscillator have been discarded to achieve steady state and $10^6$ time steps per oscillator have been used to investigate the dynamics of the system.

## RESULRS AND DISCUSSION

To gain insight into the zero-lag synchronization phenomena and its stability, we compute and analyze the time shifted cross correlation function [2, 5] between the lasers. We have chosen Figure 2 &3 for the phase-flip [5]

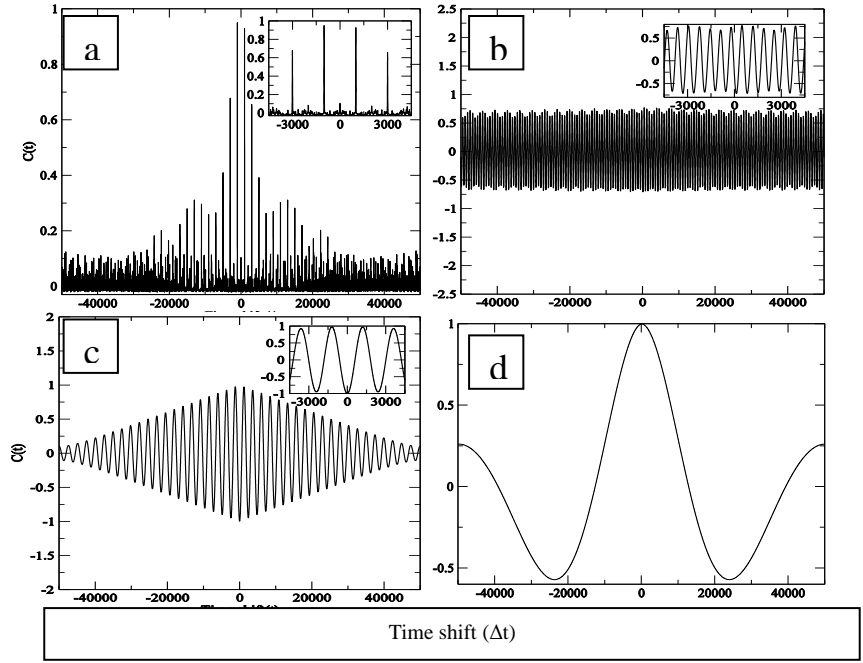

**FIGURE 2.** Time-shifted cross-correlation $C(\Delta t)$, given in [2], versus the time-shift t (in units of cavity photon lifetime), showing leading-lagging synchronization (a, b), zero-lag anti-phase (c) and in-phase synchronization (d) for a fixed time delay $\tau=14$ in the same time unit.

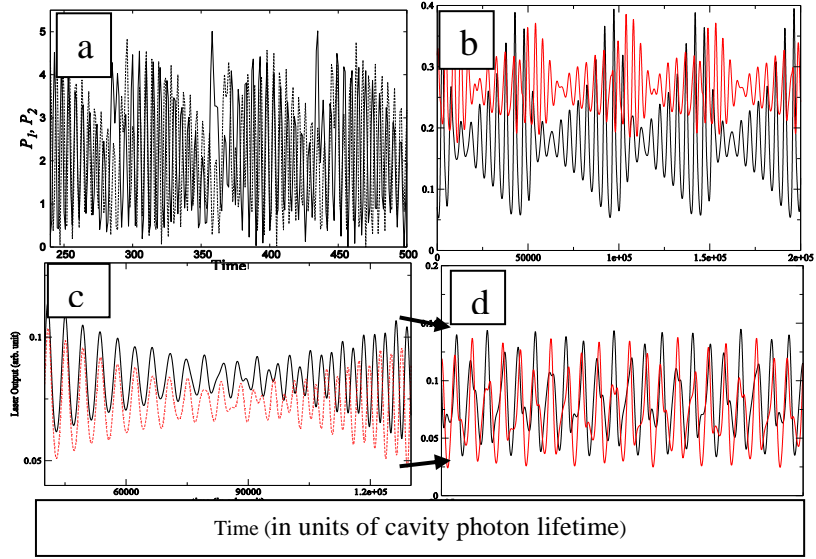

**FIGURE 3.** Plots of laser output powers, P1 and P2 versus time (in units of cavity photon life-time) for a fixed time delay τ=14, corresponding to different regimes near around the phase-flip transition [6]. (a, b) showing strange pulses, (b) coexistence of multi-attractors (d) temporal behavior of (c) extended up to long time.

In order to check the robustness of the anti-phase periodic oscillations within the window, we plot the time-shifted cross-correlation, given by the expression

$$C(\Delta t) = \frac{\langle (P_1(t) - \langle P_1(t) \rangle)(P_2(t + \Delta t) - \langle P_2(t) \rangle) \rangle}{\sqrt{\langle (P_1(t) - \langle P_1(t) \rangle)^2 \rangle \langle (P_2(t) - \langle P_2(t) \rangle)^2 \rangle}},$$

as a function of the time lag , shown in figure 2. In order to mimic the real experimental observation [1], we have chosen τ = 14 in our model. We see that the correlation is maximum at zero lag, which means that the locked anti-phase dynamics within the window is independent of the delay time. This state is independent of the delay time which is responsible for leading–lagging behavior in mutually coupled diode laser systems.

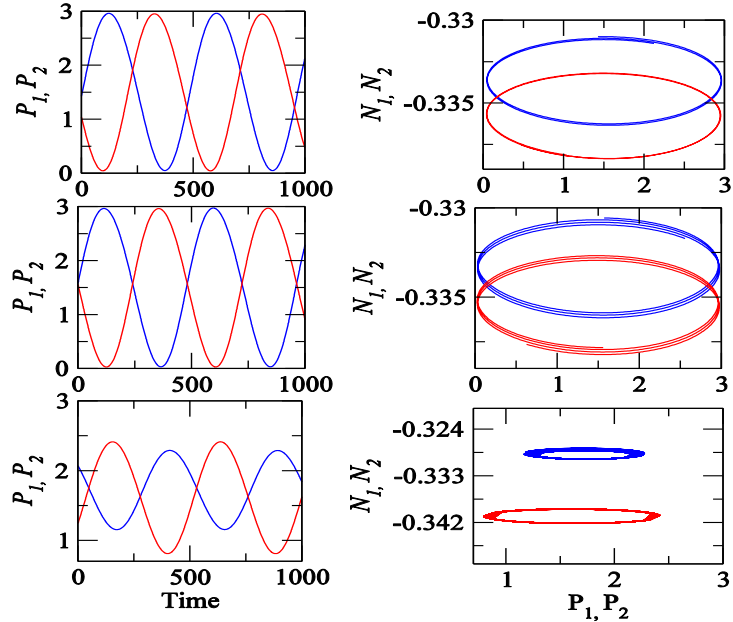

**FIGURE 4.** Plots of laser output powers P1 (blue line) and P2 (red line) versus time (in units of cavity photon lifetime) for a fixed time delay τ=14 in the same time units (left column) the out-of-phase periodic state in the transition regime; (right column) phase space plot of (left). The initial transients of $10^5$

One can easily notice that the results [see Figure 2] show the maximum correlation occur at different time for different coupling strength near the phase-flip transition regime [5]. The correlation function defined in such a way that the maximum cross correlation at a positive time difference $\Delta t_{max}$, indicates that one laser is leading another laser by the time $\Delta t_{max}$, and vice versa. The zero-lag in-phase and anti-phase synchronization in cross correlation function shown in figure 2 (c, d) which represent a maximum of -1 and + 1 at $\Delta t_{max}= 0$ (i.e. zero delay). In addition, we have systematically studied, by means of numerical simulation, the temporal collective behavior [see figure 3] near the phase-flip transition regime at various coupling strength for a fixed time delay. We have also observed periodic out-of-phase oscillation near and within the phase-flip transition regimes [see figure 4]. Numerical analysis is based on the observation that the time-shifted correlation measure unveils the signature of coexistence of multiple attractors and enhancement of stability within anti-phase amplitude-death islands [see figure 3 c & 3d]. Our finding distinguishes itself by its robustness against variations of system parameters, even in strongly coupled ensembles of oscillators

## CONCLUSION

We have been able to find the zero-lag synchronization around the regimes of phase-flip transition in delay-coupled diode lasers system without using any relay element or self-feedback mechanism. We have studied the collective behavior of two delay-coupled diode lasers, showing that optical coupling strength induced modulation of phase-amplitude coupling factor [5, 9] leads to zero-phase lag synchronization even in the presence of coupling delays. This behavior corresponds to a stable isochronous synchronization solution of the dynamics, and is possible irrespective of the distance between lasers system.

## ACKNOWLEDGMENTS

I wish to thanks the IISER Mohali, India for supporting this research and my fellowship. I would also like to thank Prof. R. Ghosh, JNU, New Delhi, India for stimulating discussions about the optical phase locking problem during my Ph. D tenure. The invaluable help and support of Prof. Kamal P. Singh hereby grateful acknowledged.

*Corresponding address: pk6965@gmail.com , pramod@iisermohali.ac.in